\documentclass[pra,preprint,showpacs,byrevtex]{revtex4}
\usepackage{graphicx}
\usepackage{amssymb}
\usepackage{amsmath}
\begin{document}

\def\cl{\centerline}
\def\bd{\begin{description}}
\def\be{\begin{enumerate}}
\def\ben{\begin{equation}}
\def\benn{\begin{equation*}}
\def\een{\end{equation}}
\def\eenn{\end{equation*}}
\def\benr{\begin{eqnarray}}
\def\eenr{\end{eqnarray}}
\def\benrr{\begin{eqnarray*}}
\def\eenrr{\end{eqnarray*}}
\def\ed{\end{description}}
\def\ee{\end{enumerate}}
\def\al{\alpha}
\def\b{\beta}
\def\bR{\bar\R}
\def\bc{\begin{center}}
\def\ec{\end{center}}
\def\dg{\dagger}
\def\d{\dot}
\def\D{\Delta}
\def\del{\delta}
\def\ep{\epsilon}
\def\g{\gamma}
\def\G{\Gamma}
\def\h{\hat}
\def\iny{\infty}
\def\La{\Longrightarrow}
\def\la{\lambda}
\def\m{\mu}
\def\n{\nu}
\def\noi{\noindent}
\def\Om{\Omega}
\def\om{\omega}
\def\p{\psi}
\def\pr{\prime}
\def\r{\ref}
\def\R{{\bf R}}
\def\ra{\rightarrow}
\def\up{\uparrow}
\def\dn{\downarrow}
\def\lr{\leftrightarrow}
\def\s{\sum_{i=1}^n}
\def\si{\sigma}
\def\Si{\Sigma}
\def\t{\tau}
\def\th{\theta}
\def\Th{\Theta}

\def\vep{\varepsilon}
\def\vp{\varphi}
\def\pa{\partial}
\def\un{\underline}
\def\ov{\overline}
\def\fr{\frac}
\def\sq{\sqrt}
\def\ot{\otimes}
\def\tf{\textbf}
\def\WW{\begin{stack}{\circle \\ W}\end{stack}}
\def\ww{\begin{stack}{\circle \\ w}\end{stack}}
\def\st{\stackrel}
\def\Ra{\Rightarrow}
\def\R{{\mathbb R}}
\def\mf{\mathbf }
\def\bi{\begin{itemize}}
\def\ei{\end{itemize}}
\def\i{\item}
\def\bt{\begin{tabular}}
\def\et{\end{tabular}}
\def\lf{\leftarrow}
\def\nn{\nonumber}
\def\va{\vartheta}
\def\wh{\widehat}
\def\vs{\vspace}
\def\Lam{\Lambda}
\def\sm{\setminus}
\def\ba{\begin{array}}
\def\ea{\end{array}}
\def\bd{\begin{description}}
\def\ed{\end{description}}
\def\lan{\langle}
\def\ran{\rangle}

\title{Multipartite entanglement in fermionic systems via a geometric measure}

\author{Behzad Lari\footnote{Electronic address:behzadlari1979@yahoo.com}, P. Durganandini\footnote{Electronic address: pdn@physics.unipune.ernet.in} and  Pramod S. Joag\footnote{Electronic address: pramod@physics.unipune.ernet.in}}
\affiliation{Department of Physics,
             University of Pune,
              Pune 411 007, India.}

\begin{abstract}

We study multipartite entanglement in a system consisting of indistinguishable fermions. Specifically, we have proposed a geometric entanglement measure for $N$ spin-$\fr{1}{2}$ fermions distributed over $2L$ modes (single particle states). The measure is defined on the $2L$ qubit space isomorphic to the Fock space for $2L$ single particle states. This entanglement measure is defined for a given partition of $2L$ modes containing $m\geq 2$ subsets. Thus this measure applies to $m\leq 2L$ partite fermionic system where $L$ is any finite number, giving the number of sites. The Hilbert spaces associated with these subsets may have different dimensions.  Further, we have defined the local quantum operations with respect to a given partition of modes. This definition is generic and unifies different ways of dividing a fermionic system into subsystems. We have shown, using a representative case, that the geometric measure is invariant under local unitaries corresponding to a given partition. We explicitly demonstrate the use of the measure to calculate multipartite entanglement in some correlated electron systems. To the best of our knowledge, there is no usable entanglement measure of $m>3$ partite fermionic systems in the literature, so that this is the first measure of multipartite entanglement for fermionic systems going beyond the bipartite and tripartite cases.
\end{abstract}

\maketitle

\section{\label{sec:Intro}INTRODUCTION}
The nonlocal correlations implied by the states of multipartite quantum systems form the basis of quantum information processing technologies like quantum teleportation and quantum computation ~\cite{Dirk,wooters, Ye}. Recently the question of understanding and using entanglement in the systems of identical and indistinguishable particles has seen a surge of interest~\cite{zanardi,indpar,wiseman, amico}. Two or more identical particles become indistinguishable when their wave functions overlap. Such a situation can arise, for example, in a quantum device based on quantum dot technology~\cite{Loss-Burk, Petta}. Here qubits are realized by the spins of the electrons in a system of quantum dots. The overlap between the electron wave functions in different dots can be varied by controlling parameters like gate voltages or magnetic field which change the tunneling amplitudes of the electrons from one dot to the other. For non-negligible overlaps, the entanglement between the qubits is then intimately connected to the electron entanglement which is essentially that of indistinguishable fermions. Entanglement is also expected to play a fundamental role in many physical phenomena like quantum phase transitions, quantum Hall effect, etc ~\cite{qpt, Hal} involving many body quantum systems.

The study of entanglement in many body quantum systems consisting of identical particles has posed challenging fundamental questions about the definition and nature of entanglement in such systems due to the inherent indistinguishability of particles ~\cite{zanardi, Eck,Gir,Fang}. For example, the (anti)symmetrization necessary  for indistinguishable (fermions) bosons already leads to quantum entanglement. However, it is known that the correlations due to symmetrization or anti-symmetrization are not by themselves a physically useful resource for quantum information and communication technologies; for example, there is no measurement we can do locally on a fermion in a localized state which is affected by the existence of identical fermions in other parts of the universe ~\cite{peres}. Most studies of entanglement in many body systems consisting of identical particles have focussed on the study of bipartite entanglement. There have been few studies on multipartite entanglement ~\cite{multi, multi2} although the study of multipartite entanglement in such systems is also an important and interesting question.

In this work, we study multipartite entanglement in a system consisting of identical particles. We use the idea due to Zanardi~\cite{zanardi} whereby the Fock space of a system of fermions is mapped to the isomorphic qubit or `mode' space. We then discuss entanglement in this `mode' space via a geometric measure. The idea is to use the Bloch representation of the state of the $m$-partite quantum system ~\cite{kim}. The measure is defined by the Euclidean norm of the $m$-partite correlation tensor in the Bloch representation. This correlation tensor contains all information of genuine $m$-partite entanglement (see section III). Such a measure was proposed earlier by one of the authors (and Hassan) for $N$ qubit and $N$ qudit pure states~\cite{ps1,ps2} and shown to satisfy most of the properties expected of a good measure. The Bloch representation of a quantum state has a natural geometric interpretation, which is why we call this measure a geometric measure ~\cite{kim}. Other geometric measures are based on the distance of the given state from the set of separable states in the Hilbert space ~\cite{ved,wei}. An important question in the context of quantum entanglement is that of locality. For indistinguishable particles distributed over `modes' (which are taken to be single particle states of particles constituting the system), local operations have meaning only in the context of partitions over modes. A quantum operation confined to a single subset in a given partition is then a local operation. We therefore define entanglement in such a system with respect to partitions and require it to be invariant under local unitaries defined with respect to a given partition. We explicitly demonstrate the use of the measure to calculate entanglement in some correlated electron systems.

 The outline of the paper is as follows: In Sec.\ref{sec:basic},  we begin by briefly reviewing some details about the isomorphism between the Fock space of a system of indistinguishable particles and the `mode' space \cite{zanardi}. This will help us in defining various quantities and also set up notation necessary for the subsequent analysis. In  Sec.\ref{sec:g-measure}, we define and construct the geometric measure. Various properties of the measure are also discussed with reference to a specific example in Sec.\ref{sec:fourmode}. In Sec.\ref{sec:appl}, we use the measure to study entanglement in the Hubbard dimer and trimer. Finally, we conclude in Sec.\ref{sec:summ}.

\section{\label{sec:basic}Mapping between Fock space and qudit space}

We deal with $N$ spin-$\fr{1}{2}$ fermions on a $L$ site lattice. The total number of available (localized) single particle states are then $2L$ in number. The fermionic Fock space in the occupation number representation has basis states of the form $|n_1n_2\ldots n_{2L}\ran\;\;(n_i=0,1\;;\;i=1,\ldots ,2L)$. We further assume that the total number of particles is conserved. This means that we only deal with subspaces of the Fock space corresponding to a fixed eigenvalue for the total number operator. We shall refer to this number super-selection rule as N-SSR. For a $N$-fermion system, we call such a subspace of the Fock space `$N$-sector' and denote it by $F_N$. The $N$-sector of a $2L$ mode system is the subspace $F_N$ of dimension $\binom{2L}{N}$ of the Fock space with the dimension of the Fock space for $2L$ single particle states $(N=0,1,2,\ldots ,2L)$ being
\ben \label{e1}
\sum_{N=0}^{2L}\binom{2L}{N}\;=\;2^{2L}
\een
Since a $2L$ qubit Hilbert space $(\mathbb{C}^{2})^{\otimes 2L}$ has exactly this dimension, it is possible to construct an isomorphism between the Fock space and the $2L$ qubit Hilbert space $(\mathbb{C}^{2})^{\otimes 2L}$ ~\cite{halmos}. The particular isomorphism we implement is
\ben  \label{e2}
|n_1n_2\ldots n_{2L}\ran\ra |n_1\ran\otimes |n_2\ran\otimes \cdots \otimes |n_{2L}\ran \;;\;n_i=0,1\;;\;i=1,2,\ldots ,2L
\een
where, in qubit space we associate $|0\ran\lr |\up\ran$ and $|1\ran\lr |\dn\ran .$ Note that the Slater rank of the Fock basis states $|n_1n_2\ldots n_{2L}\ran $  is $1$ so that these are separable states. Thus the above isomorphism maps separable basis states in Fock space to the separable basis states in qubit space. Further, the subspace structure of the Fock space namely,
\ben  \label{e3}
F_{2L}=\bigoplus _{N=0}^{2L}F_{N}
\een
is carried over to the qubit space under mapping (Eq.(\r{e2})) because each subspace of the Fock space with conserved fermion number $N$ is mapped to a subspace of the $2L$ qubit space spanned by the basis vectors with $N$ ones and $2L-N$ zeros. We can write
\ben  \label{e4}
H_{2L}=(\mathbb{C}^{2})^{\otimes 2L}=\bigoplus _{N=0}^{2L}H_{2L}(N)
\een
where $H_{2L}(N)$ is the image of $F_{N}$ in Eq.(\r{e3}) under the map given by Eq.(\r{e2}).

Next crucial step is to transfer the action of the creation and annihilation operators on Fock space to the qubit space, under the isomorphism given by Eq.(2) ~\cite{Cab1}. We need the creation and annihilation operators $a$ and $a^{\dg}$ acting on a single qubit state,
\benr \label{e5}
a|0\ran  =  0,\;\;a|1\ran  =  |0\ran \;\;\ && \nonumber \\
a^{\dg}|0\ran  =  |1\ran,\;\;a^{\dg}|1\ran  =  0 \nonumber  \\
\eenr
such that,
\benr  \label{e6}
a_i & \ra & I\otimes I\otimes\cdots\otimes \underbrace{a}_{i th\; qubit}\otimes \cdots\otimes I  \;\; \nonumber   \\
a_{i}^{\dg} & \ra & I\otimes\cdots\otimes\underbrace{a^{\dg}}_{i th\; qubit}\otimes \cdots\otimes I  \nonumber   \\
\eenr
Here $a_i$ ($a_i^{\dg}$) is the annihilation (creation) operator acting on Fock space $F_{2L},$ annihilating (creating) a fermion in $i$th mode. $I$ is the identity on single qubit space. The tensor product satisfying the correspondence in Eq.(\r{e6}) must be consistent with the anti-commutation property of the Fock space creation and annihilation operators,
\ben  \label{e7}
\{a_i,a_j^{\dg}\}=\del_{ij}\;\;\{a_i,a_j\}=0=\{a_i^{\dg},a_j^{\dg}\}
\een
This requirement leads to the following action of the tensor product operators on the $2L$ qubit states
\benr \label{e8}
(I\otimes I\otimes\cdots\otimes \underbrace{a(a^{\dg})}_{ith\;place}\otimes \cdots\otimes I)(|n_1\ran\otimes \cdots \otimes|n_i\ran \otimes \cdots \otimes|n_{2L}\ran) & = & (-1)^{\sum_{j=i+1}^{2L}n_j}   \nonumber  \\
  (|n_1\ran\otimes  \cdots \otimes\underbrace{a(a^{\dg})|n_i\ran}_{ith\; qubit} \otimes\cdots\otimes|n_{2L}\ran)   \nonumber  \\
\eenr
Here $n_i\in\{0,1\}\;;\;i\in\{1,2,\ldots ,2L\}$ and $\sum_{j=i+1}^{2L}n_j$ is evaluated $\mod{2}.$ Using Eq.(\r{e8}), it is straightforward to see that
\ben  \label{e9}
\{I\otimes I\otimes\cdots\otimes \underbrace{a(a^{\dg})}_{ith\;place}\otimes \cdots\otimes I\;,\; I\otimes I\otimes\cdots\otimes \underbrace{a(a^{\dg})}_{jth\;place}\otimes \cdots\otimes I\}
(|n_1\ran\otimes \cdots \otimes|n_i\ran \otimes \cdots \otimes|n_{j}\ran\otimes \cdots \otimes
|n_{2L}\ran)=0
\een
and
\benr  \label{e10}
\{I\otimes I\otimes\cdots\otimes \underbrace{a}_{ith\;place}\otimes \cdots\otimes I\;,\; I\otimes I\otimes\cdots\otimes \underbrace{a^{\dg}}_{jth\;place}\otimes \cdots\otimes I\}&&  \nonumber   \\
(|n_1\ran\otimes \cdots \otimes|n_i\ran \otimes \cdots \otimes|n_{j}\ran\otimes \cdots \otimes
|n_{2L}\ran) =  \underbrace{(I\otimes \cdots \otimes I)}_{2L\;factors}\del_{ij}   \nonumber   \\
\eenr
We note that the phase factors appearing in Eq.(\r{e8}) are the consequence of the conservation of the parity operator ~\cite{cir}
\ben \label{e11}
\hat{P}=\Pi_{i=1}^{2L}(1-2a_i^{\dg}a_i).
\een

Henceforth, in this paper, by `fermions' we mean spin-$\fr{1}{2}$ fermions. Further, we call a single particle state a mode. Thus two spin-$\fr{1}{2}$ fermions on two sites is a four mode system. In general, $N$ spin-$\fr{1}{2}$ fermions on $L$ sites is equivalent to $N$ fermions on $K=2L$ modes. For example, two spin-$\fr{1}{2}$ fermions on a two site lattice, $A,B$ say, generate four single particle states or modes $|A\up\ran,|A\dn\ran,|B\up\ran,|B\dn\ran$. In this work, we deal with entanglement between subsets forming a partition of a $2L$-mode fermionic system. We define the entanglement measure for any such partition of a $2L$-mode system without any restriction on the number and the size of the subsets forming the partition. These subsystems may involve different degrees of freedom, for example, we can deal with the entanglement between spins and sites or entanglement between two spins on the same site (intra-site entanglement). Or if each of the subsets partitioning the $2L$ modes comprises modes with common site label we have the entanglement between sites or the `site entanglement'. Thus all physically realizable subsystems of $N$ fermions over $2L$ single particle states can be addressed by dividing the $2L$ modes into suitable partitions, for example the `particle entanglement' defined in ~\cite{wiseman}.

 We now define the local and non-local operations on the $2L$ mode fermionic system ~\cite{zanardi,indpar,wiseman, zoz,Plas}. We do this by using the corresponding operations on the isomorphic qubit space $H_{2L}.$ We note that, due to isomorphism between $F_{2L}$ (Eq.(\r{e2})) and $H_{2L}$ (Eq.(\r{e4})), partitioning $2L$ modes is equivalent to the corresponding partitioning of the $2L$ qubit system into subsystems. Locality is defined with respect to the partition of $2L$ qubits (or, the corresponding partition of $2L$ modes) between whose subsets we are seeking entanglement. The operations on the state space of a single subset in a partition of $2L$ qubits is taken to be local. The entanglement measure defined with respect to a partition must be invariant under a unitary operation which is local with respect to that partition. We will illustrate this point later, using the geometric entanglement measure defined below. Henceforth `mode' and `qubit' are taken to be synonymous and we shall use the expression `modes' instead of `qubits'. In other words, the spaces $F_{2L}$ and $H_{2L}$ are taken to be the same.

\section{\label{sec:g-measure}GEOMETRIC MEASURE FOR ENTANGLEMENT.}
\subsection{\label{sec:def}Definition}

  We define a geometric measure for the partitions of the $2L$ mode $N$ fermion systems in pure states. Although the definition of the measure is quite general, the fermion number super-selection rule restricts the pure states to the appropriate subspace corresponding to $N$ fermions, namely $H_{2L}(N)$ (Eq.(\r{e4})). Thus a state $|0110\ran \in H_{4}(2)$ may be partitioned as $|01\ran\otimes|10\ran$ or as $|011\ran\otimes|0\ran$ etc where the definition of the tensor product is consistent with Eq.(\r{e8}). We use the Bloch representation of $N-$partite states (drawn from $H_{2L}(N)$) to get this measure ~\cite{ps1,ps2}.

 First we assume that a partition equally divides $2L$ modes into subsets, ie all subsets in the partition contain equal number of modes, say $n.$ This corresponds to the case of $H_{2L}$ divided into subspaces of dimension $d=2^n ,\;n$ being some divisor of the number of modes $2L.$ To get the entanglement measure we expand the state $\rho=|\p\ran\lan\p|$ of the system, supported in the appropriate $H_{2L}(N),$ in its Bloch representation.

  In order to give the Bloch representation of a density operator acting on the Hilbert space $\mathbb{C}^{d} \otimes \mathbb{C}^{d} \otimes \cdots \otimes \mathbb{C}^{d}$ of an $m=2L/n$-qudit quantum system, we introduce following notation ~\cite{ps2}. We use $k$, $k_i \; (i=1,2,\ldots)$ to denote a qudit chosen from $m$ qudits, so that $k$, $k_i \; (i=1,2,\ldots)$ take values in the set  $\mathcal{N}=\{1,2,\ldots,m\}$. The variables $\alpha_k \;\mbox{or} \; \alpha_{k_i}$ for a given $k$ or $k_i$ span the set of generators of $SU(d)$ group  for the $k$th or $k_i$th qudit, namely the set $\{\la_1,\la_2,\cdots,\la_{{d}^2-1}\}$ for the $k_i$th qudit. For two qudits $k_1$ and $k_2$ we define

   $$\lambda^{(k_1)}_{\alpha_{k_1}}=(I_{d}\otimes I_{d}\otimes \dots \otimes \lambda_{\alpha_{k_1}}\otimes I_{d}\otimes \dots \otimes I_{d})   $$
   $$\lambda^{(k_2)}_{\alpha_{k_2}}=(I_{d}\otimes I_{d}\otimes \dots \otimes \lambda_{\alpha_{k_2}}\otimes I_{d}\otimes \dots \otimes I_{d})  $$
   $$\lambda^{(k_1)}_{\alpha_{k_1}} \lambda^{(k_2)}_{\alpha_{k_2}}=(I_{d}\otimes I_{d}\otimes \dots \otimes \lambda_{\alpha_{k_1}}\otimes I_{d}\otimes \dots \otimes \lambda_{\alpha_{k_2}}\otimes I_{d}\otimes I_{d})   $$
where  $\lambda_{\alpha_{k_1}}$ and $\lambda_{\alpha_{k_2}}$ occur at the $k_1$th and $k_2$th places (corresponding to $k_1$th and $k_2$th qudits respectively) in the tensor product and are the $\alpha_{k_1}$th and  $\alpha_{k_2}$th generators of $SU(d),\; \alpha_{k_{1,2}}=1,2,\dots,d^2-1\;$. Then we can write
\ben \label{eq:erho}
\rho=\fr{1}{d^N} \{ I_{d}^{\otimes^m}+ \sum_{k \in \mathcal{N}}\sum_{\alpha_{k}}s_{\alpha_{k}}\lambda^{(k)}_{\alpha_{k}} +\sum_{\{k_1,k_2\}}\sum_{\alpha_{k_1}\alpha_{k_2}}t_{\alpha_{k_1}\alpha_{k_2}}\lambda^{(k_1)}_{\alpha_{k_1}} \lambda^{(k_2)}_{\alpha_{k_2}}+\cdots +$$
$$\sum_{\{k_1,k_2,\cdots,k_M\}}\sum_{\alpha_{k_1}\alpha_{k_2}\cdots \alpha_{k_M}}t_{\alpha_{k_1}\alpha_{k_2}\cdots \alpha_{k_M}}\lambda^{(k_1)}_{\alpha_{k_1}} \lambda^{(k_2)}_{\alpha_{k_2}}\cdots \lambda^{(k_M)}_{\alpha_{k_M}}+ \cdots+\sum_{\alpha_{1}\alpha_{2}\cdots \alpha_{N}}t_{\alpha_{1}\alpha_{2}\cdots \alpha_{N}}\lambda^{(1)}_{\alpha_{1}} \lambda^{(2)}_{\alpha_{2}}\cdots \lambda^{(m)}_{\alpha_{N}}\}
\een
where $\textbf{s}^{(k)}$ is a Bloch vector corresponding to $k$th qudit, $\textbf{s}^{(k)} =[s_{\alpha_{k}}]_{\alpha_{k}=1}^{d^2-1} $ which is a tensor of order one defined by
 $$s_{\alpha_{k}}=\fr{d}{2} Tr[\rho \lambda^{(k)}_{\alpha_{k}}]= \fr{d}{2} Tr[\rho_k \lambda_{\alpha_{k}}],$$ where $\rho_k$ is the reduced density matrix for the $k$th qudit. Here $\{k_1,k_2,\ldots,k_M\},\; 2 \le M \le m,$ is a subset of $\mathcal{N}$ and can be chosen in $\binom{m}{M}$  ways, contributing $\binom{m}{M}$ terms in the sum $\sum_{\{k_1,k_2,\cdots,k_M\}}$ in Eq.(\r{eq:erho}), each containing a tensor of order $M$. The total number of terms in the Bloch representation of $\rho$ is $2^m$. We denote the tensors occurring in the sum $\sum_{\{k_1,k_2,\cdots,k_M\}},\; (2 \le M \le m)$ by $\mathcal{T}^{\{k_1,k_2,\cdots,k_M\}}=[t_{\alpha_{k_1}\alpha_{k_2}\cdots \alpha_{k_M}}]$ which  are defined by

 $$t_{\alpha_{k_1}\alpha_{k_2}\dots\alpha_{k_M}}=\fr{d^M}{2^M} Tr[\rho \lambda^{(k_1)}_{\alpha_{k_1}} \lambda^{(k_2)}_{\alpha_{k_2}}\cdots \lambda^{(k_M)}_{\alpha_{k_M}}]$$

$$ =\fr{d^M}{2^M} Tr[\rho_{k_1k_2\dots k_M} (\lambda_{\alpha_{k_1}}\otimes\lambda_{\alpha_{k_2}}\otimes\dots \otimes\lambda_{\alpha_{k_M}})]   $$
where $\rho_{k_1k_2\dots k_M}$ is the reduced density matrix for the subsystem $\{k_1, k_2,\dots ,k_M\}$. Each of the  $\binom{m}{M}$ tensors of order $M$, occurring in the Bloch representation of $\rho$, contains all information about entanglement of the corresponding set of $M$ subsystems. All information on the entanglement contained in $\rho$ is coded in the tensors occurring in the Bloch representation of $\rho$. The tensor in last term in Eq.(\r{eq:erho}), we call it $\mathcal{T}^{(m)}$, contains all the  information of genuine $m$-partite entanglement. This follows from the observation that all other terms in the Bloch representation of $\rho$ (Eq.(\r{eq:erho})) correspond to subsystems comprising  $M<m$ qudits and the density operator contains all possible information about the state of the system. \\

The operators $\la_{\al_{k}}, \al_{k}= 1,2,\ldots,d^2-1$ are given by ~\cite{Mahler}

\ben
\mf{\hat{\la}}=\{\hat{u}_{12},\hat{u}_{13}, \hat{u}_{23},\ldots , \hat{v}_{12},\hat{v}_{13}, \hat{v}_{23},\ldots ,\hat{w}_1,\hat{w}_1,\ldots ,\hat{w}_{d-1}\}
\een
with
\ben
\hat{u}_{jk}=\hat{P}_{jk}+\hat{P}_{kj}   \nonumber  \\
\een
\ben
\hat{v}_{jk}=-i(\hat{P}_{jk}-\hat{P}_{kj})   \nonumber  \\
\een
\benr
w_{l}=\sqrt{\fr{2}{l(l+1)}}(\hat{P}_{11}+\cdots+\hat{P}_{ll}-l\hat{P}_{l+1,l+1})&&     \nonumber  \\
1\leq j < k\leq d \;\; ; \;\; 1\leq l\leq d-1    \nonumber    \\
\eenr
where
\ben
\hat{P}_{kl}=|k\ran\lan l|\;\;\;(k,l=1,2,\ldots ,d).      \nonumber  \\
\een
Note that each of the generators of the $SU(d)$ group $\la_i\;,\;i=1,2,\ldots ,d^2-1$ acts on a single qudit space and hence is local (see Sec.{\r{sec:basic}), apart from the phase factor contributed by their action, as given by Eq.(\r{e8}). We assume these phase factors to be absorbed in the coefficients in the expansion of the density operator $\rho.$

Let a $2L$ mode $N$ fermion system be partitioned by $m=2L/n$ subsets, each containing n modes. Then for this partition, we define the entanglement measure for a state $|\p\ran\in H_{2L}(N)$ by ~\cite{ps2}
\ben  \label{e12}
E=||\t||-||\t||_{sep}
\een
 where
 \ben   \label{e13}
 ||\t||=\sqrt{\sum_{\al_1\cdots\al_m =1}^{d^2-1}t_{\al_1\cdots\al_m}^{2}}
 \een
 and $||\t||_{sep}$ is $||\t||$ for separable (product) $m$ qudit state

 \ben  \label{e14}
 ||\t||_{sep}=\left(\fr{d(d-1)}{2}\right)^{m/2}
 \een

\subsection{\label{sec:unequal} Entanglement in partitions with unequal subsets}

  We can also generalize the definition of the entanglement to the case where the corresponding qubit subsystems have unequal dimensions. We discuss the simplest case of bi-partite entanglement with partitions having unequal dimensions, say $d_1$ and $d_2$. In this case, the definition of the geometric entanglement measure generalizes to
\ben \label{37}
E=||\t||-||\t||_{sep}
\een
 where
 \ben   \label{e38}
 ||\t||=\sqrt{\sum_{i=1}^{d_1^2-1}\sum_{j=1}^{d_2^2-1}t_{ij}^{2}}
 \een
 where
 \ben \label{e39}
 t_{ij}=\left(\fr{d_1}{2}\right)\left(\fr{d_2}{2}\right)\lan\p|\h{\la}_i\ot\h{\la}_j |\p\ran=\left(\fr{d_1}{2}\right)\left(\fr{d_2}{2}\right)K_{ij}.
 \een
 Here $\h{\la}_i \;(i=1,\ldots ,d_1^2-1)$ and $\h{\la}_j \; (j=1,\ldots ,d_2^2-1)$ are the generators of $SU(d_1)$ and $SU(d_2)$ respectively. $||\t_{sep}||$ is given by
\ben \label{e40}
||\t_{sep}||^2 = ||\mf{s}^{(1)}||.||\mf{s}^{(2)}|| = \left(\fr{d_1(d_1-1)}{2}\right)\left(\fr{d_2(d_2-1)}{2}\right).
\een
Here $||\mf{s}^{(1)}||$ and $||\mf{s}^{(2)}||$ are the norms of the Bloch vectors of the reduced density operators for each subsystem.

It is straightforward to extend these definitions to partitions containing more than two subsets.

\section{\label{sec:fourmode}Entanglement in a four mode system}

With the entanglement measure defined as above, we give an example wherein we can compute the entanglement for different partitions and  illustrate our comments on local and non-local operations. We also compute the upper bounds on the entanglement.

\subsection{\label{sec:l-nlocal}Local and Non-local operations}
 Consider a four mode system and the normalized state $|\p\ran\in H_4(2)$ defined as
\ben  \label{e15}
|\p\ran=\fr{1}{\sqrt{6}}\{ i\al |1100\ran + |1001\ran + |0110\ran + |0011\ran + \b |0101\ran + |1010\ran \}\;\;\al^2 + \b^2=2
\een
where $\al,$ $\b$ are real. Note that $|\p\ran$ can be treated as a member of the Fock space $F_4 (2)$ with the kets appearing in it being its basis states. Consider the evolution of the system in state $|\p\ran \in F_4(2)$ via the Hamiltonian
\ben  \label{e16}
H = f(a_1^{\dg}a_4+a_4^{\dg}a_1)+q\h{n}_1\h{n}_2 + \G \h{n}_1 + \g \h{n}_3 + \eta(a_1^{\dg}a_2+a_2^{\dg}a_1)
\een
acting on $F_4 .$ Here $f$ term is the interaction between two modes on different sites (inter-site interaction), $\eta$ term is the interaction between two modes on the same site (intra-site interaction). $\G$ and $\g$ correspond to single mode on site $A$ and $B$ respectively. q term involves number operators $\h{n}_i = a_i^{\dg}a_i\;;\;i=1,2$ for first two modes, on $A$ site. We have included all the different kinds of typical interactions encountered in condensed matter systems, respecting number super-selection rule. After an infinitesimal unitary evolution via this Hamiltonian, the state $|\p\ran$ evolves to
\ben   \label{f1}
|\p^{\pr}\ran=|\p\ran-i\ep H|\p\ran
\een
By employing the mapping of annihilation and creation operators in Eq.(\r{e6}) and Eq.(\r{e8}) and that of Fock space basis states in Eq.(\r{e2}), we get, for $|\p^{\pr}\ran$
\benr   \label{e17}
|\p^{\pr}\ran &=& \fr{1}{\sqrt{6}}\{(i\al+i\ep f+\al q\ep)|1100\ran + (1-i\G\ep -i\ep\eta\b)|1001\ran + (1-i\g\ep -i\ep\eta)|0110\ran +     \nonumber   \\
 && +  (1-i\ep f -i\ep\g)|0011\ran
+ (\b-\ep f\b-i\ep\eta)|0101\ran + (1+i\ep f - i\ep\G-i\ep\g - i\ep\eta)|1010\ran    \nonumber   \\
\eenr

  Now we find the entanglement for different partitions of this four mode system, using the geometric entanglement measure, Eq.(\r{e12}). We first partition four modes into four subsets, each containing one mode. This case gives genuine entanglement between four modes, which is more general than only the bipartite entanglement considered in the literature. For this case $d=2,$ so that $||\t||_{sep}=1$ and we get, for the genuine four mode entanglement,
 \ben  \label{e18}
 E=||\t||-1
 \een
 where
 \ben   \label{e19}
 ||\t||=\sqrt{\sum_{i,j,k, l=1}^{3}t_{ijkl}^2}
 \een
 with
 \ben  \label{e20}
 t_{ijkl}=Tr[\rho\;\si_i\ot\si_j\ot\si_k\ot\si_l]=\lan\p|\si_i\ot\si_j\ot\si_k\ot\si_l|\p\ran .
 \een
 where $\{\si_i\}\;i=0,1,2,3$ are the generators of the $SU(2)$ group (Pauli operators). The resulting entanglement in $|\p^{\pr}\ran$ is
 \benr  \label{e21}
 E_g(|\p^{\pr}\ran) & = & \fr{1}{6}\left(-6+\sq{88+64\al^2+32\b+64\b^2+10\al^2\b^2+\b^4}\right)  \nonumber  \\
 & & -\fr{4\left(4f\al-2q\al(1+\b)+f\al\b(\al^2-\b^2)+4\al\eta(1+\b)\right)\ep}{\left(-6+\sq{88+64\al^2+32\b+64\b^2+10\al^2\b^2+\b^4}\right)}+O[\ep^2] \nonumber \\
 \eenr
 where the first term gives the entanglement $E(|\p\ran)$ for the state $\psi$ as defined in Eq.(\r{e15}). For this partition, the operations on a single mode are the only local operations, while all others are non-local. Therefore, the terms $\G \h{n}_1$ and $\g \h{n}_3$ are the only local interactions. Therefore, we expect that the four mode genuine entanglement should not depend on $\G$ or $\g$ to the first order in $\ep,$ which is the case, as seen from Eq.(\r{e21}).

   Next, we consider the partition consisting of two subsets, each containing two modes on each site, $\{A\up ,A\dn\}$ and $\{B\up ,B\dn\}$ (site partition). Thus we have two subsystems with $d=4$ corresponding to a $SU(4)\ot SU(4)$ qudit system. Further, $n=2$ giving $m=(2L/n)=2$ so that the geometric entanglement is

 \ben \label{e22}
 E_s(|\p^{\pr}\ran)=||\t||-6
 \een
 where
 \ben
 ||\t||=4\sq{\sum_{j,k=1}^{15}K_{jk}^{2}}  \nonumber  \\
 \een
 with
 \ben
 K_{jk}=\lan\p|\h{\la}_j\ot\h{\la}_k|\p\ran  \nonumber   \\
 \een
where $\h{\la}_j \;;\;j=1,\ldots ,15$ are the generators of $SU(4).$ The entanglement of $|\p^{\pr}\ran$ in Eq.(\r{e17}) is then given by
\benr \label{e23}
E_s(|\p^{\pr}\ran)&=&\fr{1}{3}\left(-18+\sq{208+136\al^2+9\al^4-32\b+104\b^2+34\al^2\b^2+9\b^4}\right)   \nonumber  \\
&& -\fr{16(-f\al+f\al\b(\al^2-\b^2-2))\ep}{3(\sq{208+136\al^2+9\al^4-32\b+104\b^2+34\al^2\b^2+9\b^4})}+O[\ep^2]   \nonumber  \\
\eenr

According to the `site partition', in addition to the operations on single modes, the operations on the pair of modes having the same site label are also local. Therefore, the resulting entanglement cannot change under the intra-site operations in the Hamiltonian, namely the $q$ term, the $\eta$ term and as before, $\G$ and $\g$ terms. Thus, to the first order in $\ep ,$ the entanglement is expected to depend only on the non-local part of the Hamiltonian, that is, on the $f$ parameter. From Eq.(\r{e23}) we see that this is the case. The inter-site entanglement quantified using Von-Neumann entropy has been reported earlier~\cite{zanardi}.

Thus we see that the geometric measure has the capability to quantify the genuine multi-mode fermionic entanglement as against the mainly bipartite entanglement reported in the literature. Also, the geometric entanglement measure, for the given partition of modes, shows the correct behavior under local and non-local unitary operations.

\subsection{\label{sec:bound}Upper bound for the geometric measure in a four mode system.}

We discuss here some upper bounds that one can find for the entanglement for the four mode system and compare with existing results
obtained from other measures. We find that the general state which leads to a maximum inter-site entanglement computed via the geometric measure also leads to a maximum for the von-Neumann entropy.

We treat the entanglement for the given partition to be the function of the coefficients of the general state $|\p\ran\in H_4(2),$ namely,
\ben \label{e24}
|\p\ran=\sum_{k=3,5,6,9,10,12}c_k |k\ran
\een
where each ket is labeled by the (four bit) binary representation of $k.$ We then maximize $E(|\p\ran)$ with respect to the coefficients $c_k $.

 For the site partition (see above) the entanglement (Eq.(\r{e22}) and two equations following it) is a function of the coefficients $c_k$ in state
$|\p\ran\in H_4(2)$ given in Eq.(\r{e24}). We find that the the maximum value of the entanglement is given by
\ben
E_{max}=1.74593
\een
The Schmidt canonical form (with respect to the $H_{4}(2)$ basis) of $|\p_{max}\ran$ corresponding to this $E_{max}$ is
\ben \label{e27}
 |\p_{max}\ran=\fr{1}{2}\left(\sum_{k=3,5,10,12}|k\ran\right)
 \een
 or,
 \ben  \label{e28}
 |\p_{max}\ran=\fr{1}{2}\left(\sum_{k=3,6,9,12}|k\ran\right)
 \een
 where each ket is labeled by the four bit binary representation of $k.$ Also, both these forms lead to the von-Neumann entropy $=2$ which is the maximum  possible~\cite{Gir}.

For the four mode entanglement (Eq.(\r{e18})) we find that entanglement is maximum for the state given by
\begin{eqnarray}
&&c_3=0.06701-0.33751i\;\; c_5=0.16442-0.51281i\;\;c_6=-0.1006+0.2854i    \nn   \\
&&c_9=0.29967-0.04206i\;\;c_{10}=-0.53522+0.05955i\;\;c_{12}=-0.34410-0.00118i   \nn   \\
\end{eqnarray}
and the maximum value is found to be
\ben
E_{max}=2
\een
However, we do not have any entanglement measure to compare with the geometric measure. We also note that, for four modes, no canonical form like the Schmidt or Acin canonical form (for two and three modes respectively) is available.

\section{\label{sec:appl}APPLICATIONS}
We now consider some correlated fermionic lattice models and discuss multi-mode entanglement in these models using the geometric
measure.

\subsection{\label{sec:Hdimer}Hubbard dimer.}

 The Hubbard dimer model is a simple model for a number of physical systems, including the electrons in a $H_2$ molecule, double quantum dots, etc~\cite{zanardi,wiseman,hdimer}. The Hamiltonian can be written as
\ben \label{e29}
H\;=\;-t\sum_{\si=\up,\dn}\left(c^{\dg}_{A\si}c_{B\si}+c^{\dg}_{B\si}c_{A\si}\right)+U\sum_{j=A,B}\h{n}_{j\up}\h{n}_{j\dn}
\een
where $A,$ $B$ are the site labels and $\up$ and $\dn$ are spin labels. $t$ is the hopping coefficient measuring hopping between two sites while conserving spin and $U$ quantifies Coulomb interaction between fermions on the same site. By varying $\left(\fr{U}{4t}\right)$ we can vary the relative contributions of hopping and Coulomb mechanisms.

The ground state of the system at zero temperature can be easily obtained as ~\cite{zanardi,wiseman}
\ben \label{e30}
|\p_0\ran=N\h{G}_0 |vac\ran
\een
where $N=\lan \p_0|\p_0\ran^{-1/2}$ is the normalization factor and
\ben \label{e31}
\h{G}_0=c^{\dg}_{A\up}c^{\dg}_{A\dn}+c^{\dg}_{B\up}c^{\dg}_{B\dn}+\al\left(\fr{U}{4t}\right)\left(c^{\dg}_{A\up}c^{\dg}_{B\dn}-c^{\dg}_{A\dn}c^{\dg}_{B\up}\right)
\een
with $\al(x)=x+\sq{1+x^2} .$ By mapping to $H_{2L}$ via Eq.(\r{e2}) we get,
\ben \label{e32}
|vac\ran\ra|0\ran\ot|0\ran\ot|0\ran\ot|0\ran=|0000\ran
\een
while mapping between the operators, (Eq.(\r{e6}) and Eq.(\r{e8})) gives
\benr \label{e33}
c^{\dg}_{A\up}\ra a^{\dg}\ot I_2\ot I_2\ot I_2\;\;\;c^{\dg}_{A\dn}\ra I_2\ot a^{\dg}\ot I_2\ot I_2 &&   \nonumber    \\
c^{\dg}_{B\up}\ra  I_2\ot I_2\ot a^{\dg}\ot I_2\;\;\;c^{\dg}_{B\dn}\ra I_2\ot I_2\ot I_2\ot a^{\dg}      \nonumber    \\
\eenr
The normalized ground state can be expressed in the qubit space as
\ben  \label{e34}
|\p_0\ran=\fr{-1}{\sq{2(1+\al^2)}}\left\{|1100\ran+|0011\ran+\al |1001\ran-\al |0110\ran\right\}
\een

 The four-partite entanglement in any state $|\psi\ran$ can be calculated by using Eqs.((\r{e18}),(\r{e19}),(\r{e20})) as,
\ben \label{e35}
 E_g=||\t||-1
 \een
 where
 \ben   \label{e-fourp}
 ||\t||=\sq{\sum_{i,j,k,l=1}^{3}t_{ijkl}^{2}}
 \een
 with
 \ben  \label{e-fourp1}
 t_{ijkl}=\lan\p|\si_i\ot \si_j\ot \si_k\ot \si_l|\p\ran .
 \een
The ground state entanglement can be then calculated to be
\ben  \label{e-fourp2}
E_g=\fr{3}{(1+\al^2)}\sq{1+\fr{2}{9}\al^2+\al^4}-1
\een
We plot the four-partite entanglement as a function of $U$ and $t$ (Fig.1(a)) and as a function of $\al$ (Fig.1(b)). The entanglement is seen to  monotonically increase as a function of $\alpha$, saturating at large values of $\alpha$ to the maximum value $2$. The saturation to the maximum value can be obtained either for very large values of $U$ or very small values of $t$. We can interpret this result in the following way: since the total particle number is fixed to be $2$, the four mode entanglement essentially measures the correlations between the spins. The entanglement increases as a function of $\alpha$ because the spin correlations increase with $\alpha$.

\begin{figure}[!ht]
\begin{center}

\includegraphics[width=7cm,height=5cm]{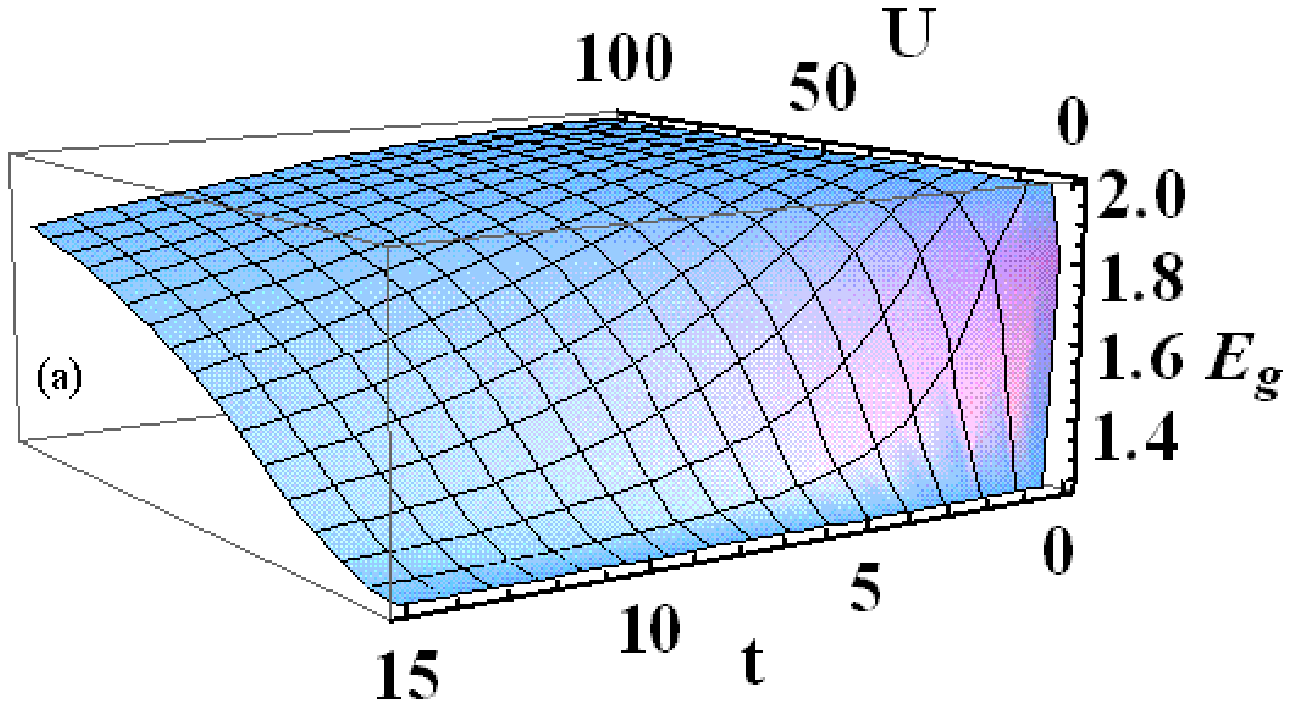}
\includegraphics[width=7cm,height=5cm]{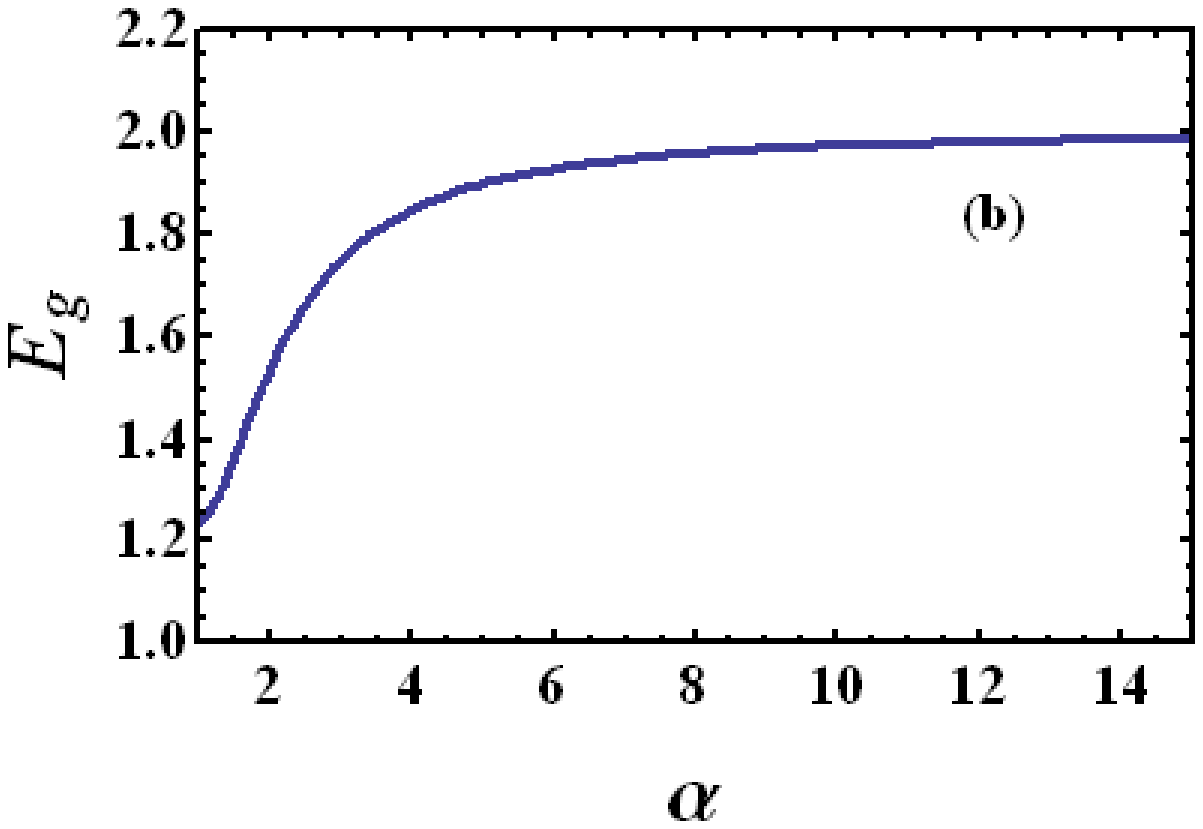}

 Fig. 1\,\,\,\,\, Four-partite entanglement for the Hubbard dimer (at half filling) as a function of (a)(color online) $U$ and $t$ (both in energy units) and (b) as a function of $\al$ ($\al\ge1$).

\end{center}
\end{figure}

We can also calculate the bipartite entanglement between sites $A$ and $B$ using the geometric measure Eqs.(\r{e12},\r{e13},\r{e14}) considering the partitions to be $\{\{A\up \; A\dn\} ; \{B\up \; B\dn\}\} $
 \ben  \label{e48}
 E_s =||\t||-||\t||_{sep}
 \een
 where
 \ben  \label{e49}
 ||\t||=\sq{\sum_{i,j= 1}^{15} t_{ij}^2}; \,\,\,\,\,  t_{ij}=\left(\fr{d}{2}\right)^2 \lan\p|\h{\la}_i\ot \h{\la}_j |\p\ran
 \een
and $||\t||_{sep}=\left(\fr{d(d-1)}{2}\right)^{m/2}$.
Here $\h{\la}$s are the generators of $SU(4),$ there are $m=2$ partitions (same as the number of sites) and each partition has dimension $d=4.$ This leads to an inter-site entanglement of the form
\ben \label{e36}
E_{s}=\fr{2}{(1+\al^2)}\sq{13\al^4+34\al^2+13}  - 6
\een
The bi-partite entanglement between sites $A$ and $B$ was calculated earlier using the von-Neumann entropy \cite{zanardi}
\ben \label{e55}
E_{VN}=\fr{1}{(1+\al^2)}\left\{\log_2\left[2(1+\al^2)\right]-\al^2\log_2\left[\fr{\al^2}{2(1+\al^2)}\right]\right\}   \\
\een
We plot the inter-site entanglement (the von-Neumann entropy is also plotted for comparison) as a function of $\al$ in Fig.2. It is seen that both measures show qualitatively similar behavior, i.e, a monotonically decreasing entanglement as a function of $\alpha$ saturating at very large values of $\al$. The entanglement between the sites $A$ and $B$ decreases as a function of $\alpha$ because  with increasing on-site repulsion $U$, the four dimensional local state space at each site gets reduced to a two dimensional local state space~\cite{zanardi} due to a suppression of charge fluctuations or in other words, as $\al \to \iny$ the $SU(4)\ot SU(4)$ partition goes over to a $SU(2)\ot SU(2)$ partition. We have explicitly checked that the entanglement obtained in the $\al \to \iny$ limit matches with that obtained for the $SU(2)\ot SU(2)$ partition.\\
\begin{figure}[!ht]

\includegraphics[width=7cm,height=5cm]{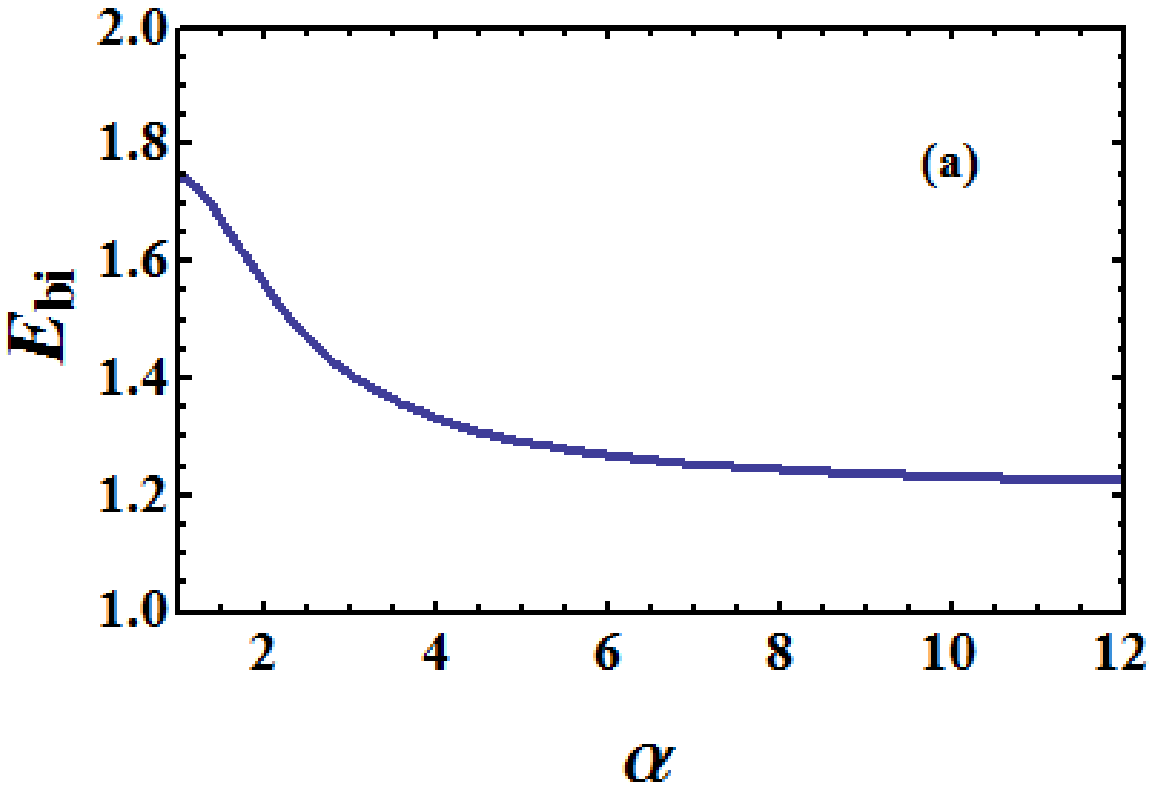}
\includegraphics[width=7cm,height=5cm]{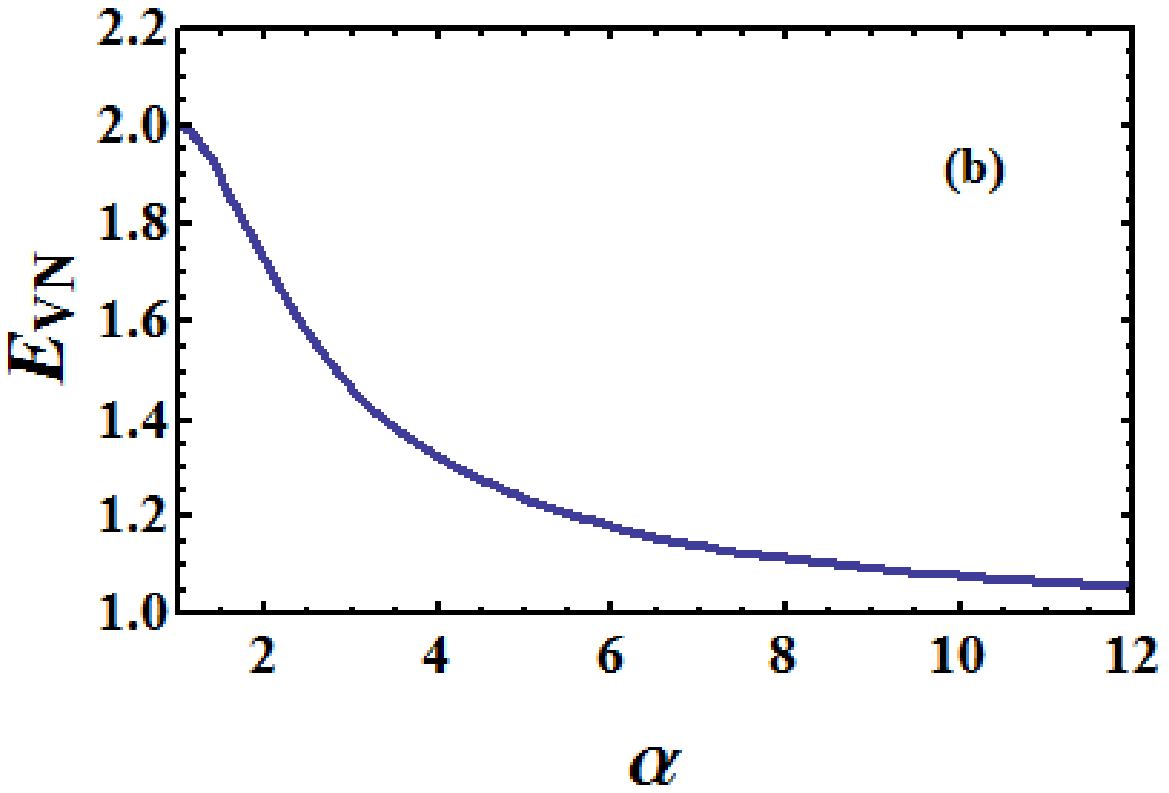}

Fig.2 (a) The bi-partite entanglement between sites A and B calculated with the geometric measure as a function of $\al$ ($\al\ge1$) for the Hubbard dimer at half filling. (b) The corresponding von-Neumann entropy as a function of $\al$ ($\al\ge1$).


\end{figure}

We can also discuss bi-partite entanglement with unequal partitioning. Consider four modes partitioned into two subsets containing one and three modes respectively. This would correspond for example to one observer controlling a register which measures the occupancy of the spin up at site $A$ and the other observer controlling a register which measures the occupancy of the spin down at site $A$ as well as the occupancy at site $B$. The bi-partite entanglement in this can be obtained as discussed in Sec.(\ref{sec:unequal}). In the present case, the two partitions have dimension as $d_1=2$ and $d_2=8$ respectively. This gives, for the entanglement,
\ben  \label{e41}
E=4\sq{\sum_{j=1}^{3}\sum_{k=1}^{63}K_{jk}^2}-\sq{28}
\een
or, using $|\psi\ran$ in Eq.(\r{e34}),
\ben \label{e42}
E=1.6367
\een
 Interestingly, the entanglement is independent of $\al$. It turns out that this is the maximum value possible for the entanglement (Eq.(\r{e41}))( we checked this by maximizing the entanglement given by Eq.(\r{e41}) as a function of the coefficients in the general state Eq.(\r{e24}). We have also checked that von-Neumann entropy in this case is also independent of $\al$ and has the maximum possible value, i.e. $2$.

\subsection{ Three electrons on three sites.}

We next consider the Hubbard trimer, i.e, electrons on three sites with the sites $A,B,C$  with periodic boundary conditions. Ignoring the chemical potential , the Hamiltonian is
 \ben   \label{e43}
 H(\beta)=-t\left[\sum_{j=A,B,C}(c_{j\up}^{\dg}c_{j+1\up}+c_{j\dn}^{\dg}c_{j+1\dn}+h.c.)-\beta \sum_{j=A,B,C}\h{n}_{j\up}\h{n}_{j\dn}\right]
 \een
 where $t>0$ is the hopping parameter and $\beta =\frac{U}{t}$. We begin our analysis as earlier by mapping from the fermionic to the six qubit space corresponding to $(A\up \; A\dn \; B\up \; B\dn \; C\up \; C\dn )$. In the qubit space, the basis respecting number super-selection rule is given by
\ben  \label{e54}
\{|k\ran\}\;\; k=7,11,13,14,19,21,22,25,26,28,35,37,38,41,42,44,49,50,52,56
\een
where the twenty basis states are labeled by the six bit binary representation of the $k$ values. We numerically diagonalize the Hamiltonian $H(\beta)$ in Eq.(\r{e43}) in the basis given by Eq.(\r{e54}) at different values of $\beta$ as $\beta$ increases from zero. The total spin quantum number $S$ and the $z$ component of the total spin $S_Z$ commute with the Hamiltonian and can therefore be used as good quantum numbers to characterize the states. The ground state has total $S$ value $=1/2$. The triangular geometry of the three site model (with periodic boundary conditions) also leads to an additional symmetry under reflection about one of the medians of the triangle. This leads to a two fold degeneracy for the ground state (for a fixed $S_Z$ value). Since these symmetries are preserved even in the presence of the interaction $U$, the ground state remains two-fold degenerate for all $\beta$ values.

The entanglement in any state can be calculated in a similar manner as shown previously (Sec.\r{sec:Hdimer}).
We now show the results of our calculations for the entanglement in one of the ground states.

The six-mode entanglement as a function of the interaction parameter $\beta$ is shown in Fig.3a while the tripartite entanglement (between the sites $A, B, C$) is shown in Fig.3b as a function of $\beta$.  The tripartite entanglement between the sites $A,B,C$ is seen to decrease with increasing $\beta$ - we interpret this result  in a similar way as that for the dimer as due to the fact that the local state space at each site decreases with increasing $\beta$.  We find that the six-mode entanglement increases with $\beta$, saturating at large values, however, the behavior is not monotonic. Such non- monotonic behavior as a function of $\beta$ is also shown by the bipartite entanglement between sites $A$ and $ BC$ calculated using the geometric measure as well as the von-Neumann entropy (Figs.3c and 3d).

      We can understand the non-monotonic behavior in the following way: the bi-partite entanglement measures the entanglement between the the sites $A$ and the sites $BC$. With small increase in $\beta$ from zero, there is an increase in the spin fluctuations between $B$ and $C$ which shows up as an initial increase in the entanglement between $A$ and $BC$. However, with further increase in $\beta$, the charge fluctuations get completely suppressed leading to an asymptotic behavior similar to that for the dimer. One can understand the non-monotonic behavior of the six-mode entanglement in a similar way. For small $\beta$, there is a larger correlation between two sites which leads to a decrease in the overall entanglement, however with further increase in $\beta$, the contribution to the entanglement is solely due to spin fluctuations which increases with $\beta$ leading to the observed increase in the entanglement as well.

\begin{figure}[!ht]

\includegraphics[width=7cm,height=5cm]{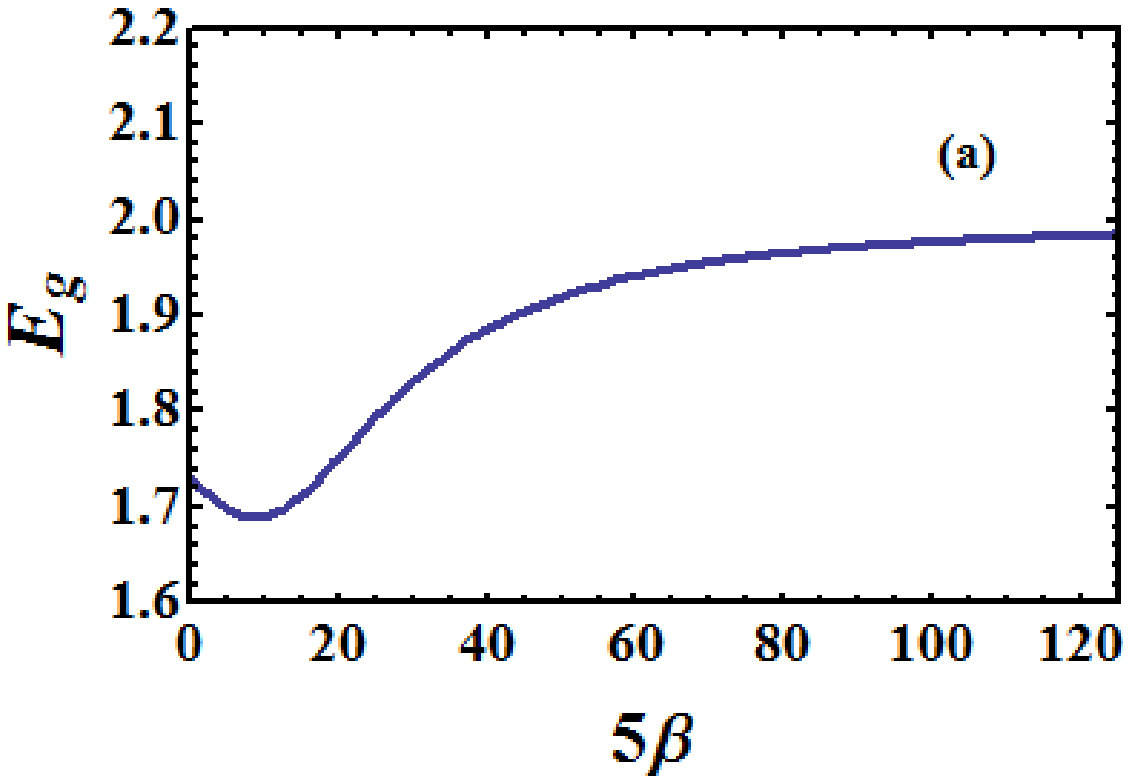}
\includegraphics[width=7cm,height=5cm]{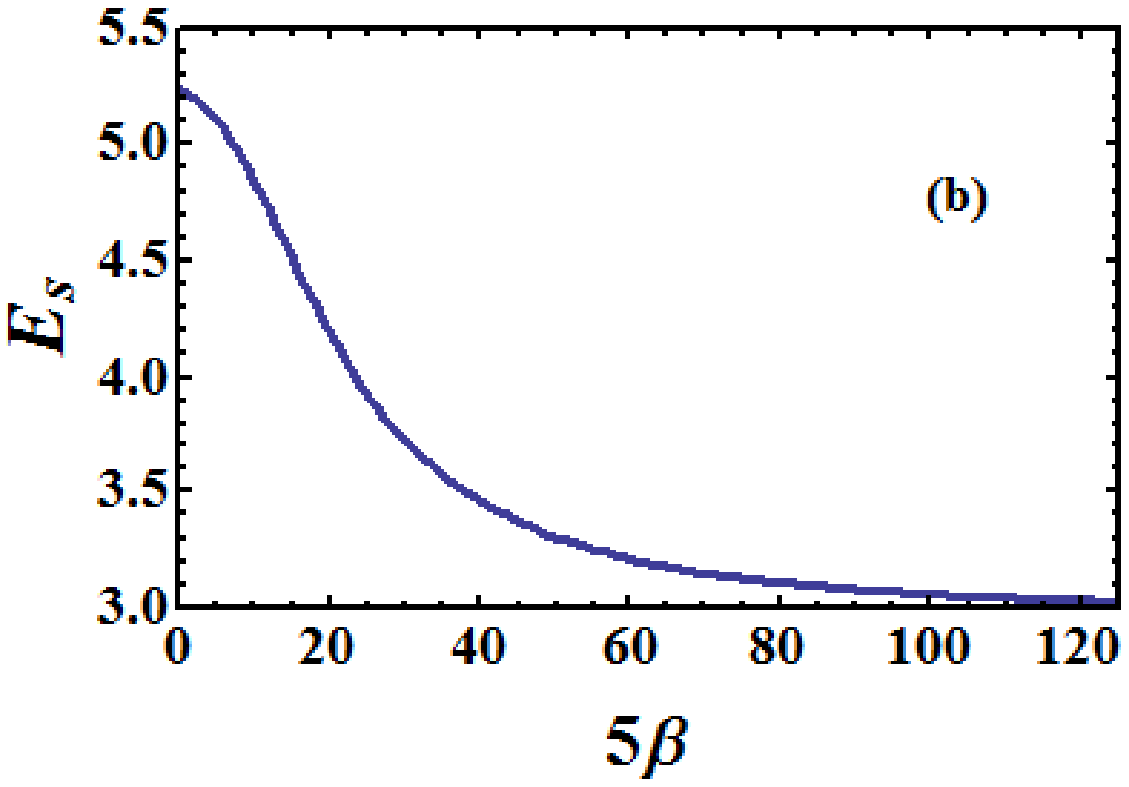}
\includegraphics[width=7cm,height=5cm]{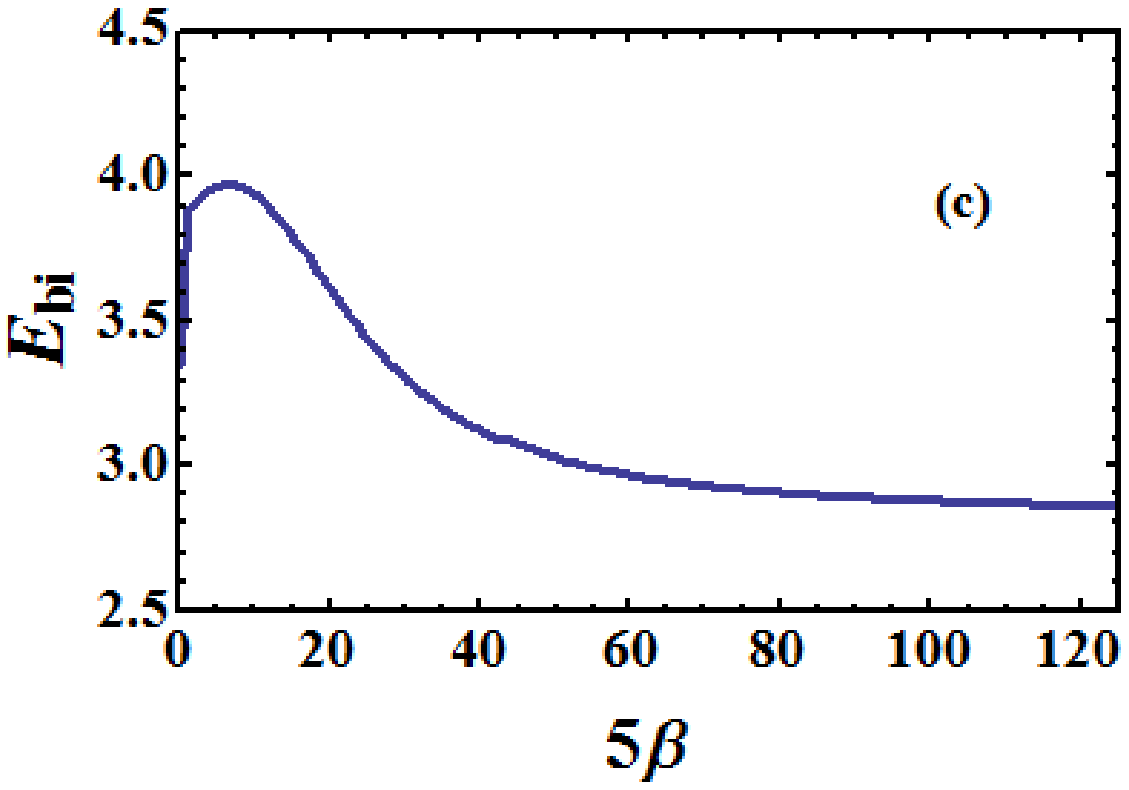}
\includegraphics[width=7cm,height=5cm]{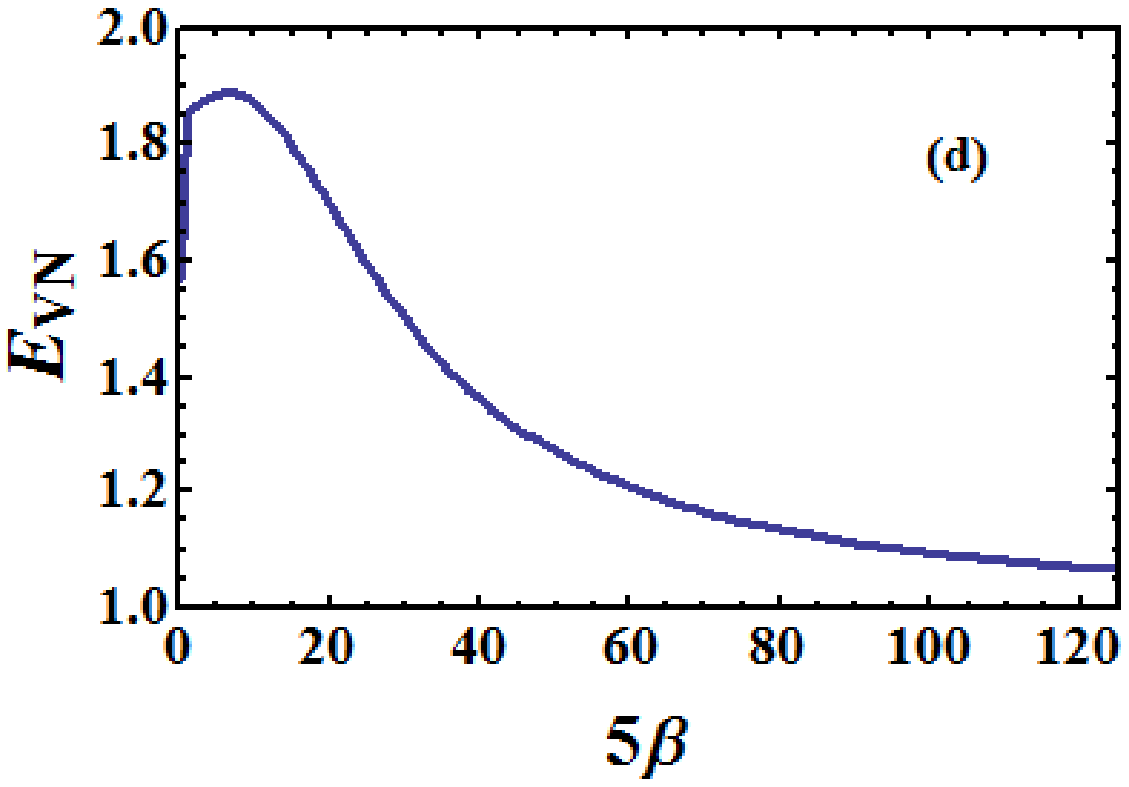}

Fig.3 (a) Six-partite entanglement for the Hubbard trimer as a function of $\beta$ at half filling, (b) Tripartite (site) entanglement as a function of $\beta$ for the Hubbard trimer at half filling, (c) The bipartite entanglement between site A and sites BC (see text) as a function of $\beta$ for the Hubbard trimer at half filling and (d) The corresponding von-Neumann entropy as a function of $\beta$ for the Hubbard trimer at half filling.

\end{figure}

We have also calculated the upper bounds for the entanglements in various partitions in the same way as in Sec.\ref{sec:fourmode}. These are $E_g^{max}=4.42218,$ $E_{bi}^{max}=4.15105,$ $E_s^{max}=6.08767.$

\section{\label{sec:summ}SUMMARY}

We have proposed a multipartite entanglement measure for $N$ fermions distributed over $2L$ modes (single particle states). The measure is defined on the $2L$ qubit space isomorphic to the Fock space for $2L$ single particle states. The entanglement measure is defined for a given partition of $2L$ modes containing $m\geq 2$ subsets, using the Euclidean norm of the $m$-partite correlation tensor in the Bloch representation of the corresponding multi-mode state, viewed as a $m$-partite state (see sec. IV and V). The Hilbert spaces associated with these subsets may have different dimensions. The quantum operations confined to a subset of a given partition are local operations. This way of defining entanglement and local operations gives us the flexibility to deal with entanglement and its dynamics in various physical situations governed by different Hamiltonians.  We note that the concept of locality for indistinguishable fermions is distinct from that for distinguishable particles. In the latter case, locality applies to spatially separated subsystems. However, spatially separated fermions become distinguishable. We have shown, using a representative case, that the geometric measure is invariant under local unitaries corresponding to a given partition. As an application, we have also considered some correlated electron systems and demonstrated the use of the multipartite measure in these systems. In particular, we have calculated the multipartite entanglement in the Hubbard dimer and trimer (at half filling). We find that the bipartite entanglement between sites as computed with the geometric measure has a qualitatively similar behavior as a function of the interaction as that of the conventional von-Neumann entropy. We have also calculated the four(six)- partite and the two (three) site entanglement for the Hubbard dimer(trimer). We find that the multi-partite entanglement gives complementary information to that of the site entanglement in both the cases.

  Although the entanglement measures given in this paper have been mainly applied to the study of the multipartite entanglement structure in Hubbard dimers and trimers, these measures are completely general and can be  applied to other fermionic systems. We have also shown ~\cite{ps1,ps2} that, viewed as a measure on qubit space, this measure has most of the properties required of a good entanglement measure, including monotonicity. To the best of our knowledge, this is the first measure of multipartite entanglement in fermionic systems going beyond the bipartite and even the tripartite case. Further, in this paper we have restricted to applications involving ground states of $2L$ mode systems with $2L\leq 6$. It is straightforward to extend the calculations for large number of modes ($2L>6$) except for the length of the computation. We have shown earlier that ~\cite{ps1, ps2} for antisymmetric states, the computational complexity of computing the measure goes polynomially with the number of parts of the system (here, the number of partitions of $2L$ modes). It would be interesting to extend these calculations to larger system sizes where new and interesting results might be expected - we plan to do this in a separate study.

\acknowledgments{It is a pleasure to thank R.Shankar, V.Subramanian, Sibasish Ghosh , Sandeep Goyal, Ali Saif M. Hassan and Ali Ahanj for useful discussions. We thank Guruprasad Kar and Prof. R.Simon for encouragement. PD and PSJ also acknowledge support from their respective BCUD, Pune University grants.}

\end{document}